\documentclass[twocolumn,aps,prl,showpacs,preprintnumbers,amsmath,amssymb]{revtex4}
\usepackage{graphicx}
\usepackage{dcolumn}
\usepackage{bm}
\begin{document}
\title{Artificial Seismic Shadow Zone by Acoustic Metamaterials}
\author{Sang-Hoon  \surname{Kim}$^{a}$} \email{shkim@mmu.ac.kr }
\author{Mukunda P. \surname{Das}$^{b}$}\email{mukunda.das@anu.edu.au}
\affiliation{
$^a$Division of Marine Engineering, Mokpo National Maritime University,
Mokpo 530-729, R. O. Korea
\\
$^b$Department of Theoretical Physics, RSPE,
The Australian National University, Canberra, ACT 0200, Australia
}
\begin{abstract}
We developed a new method of earthquakeproof engineering
to create an artificial seismic shadow zone using acoustic metamaterials.
By designing huge empty boxes with a few side-holes corresponding to the resonance frequencies
of seismic waves and burying them around the buildings that we want to protect,
 the velocity of the seismic wave becomes imaginary.
The meta-barrier composed of many meta-boxes attenuates the seismic waves,
which reduces the amplitude of the wave exponentially by dissipating the seismic energy.
This is a mechanical method of converting the seismic energy into sound and heat.
We estimated the sound level generated from a seismic wave.
This method of area protection differs from the point protection
of conventional seismic design, including the traditional cloaking method.
The meta-barrier creates a seismic shadow zone, protecting all the buildings within the zone.
The seismic shadow zone is tested by computer simulation and compared with a normal barrier.
\end{abstract}
\date{\today}
  \maketitle


An earthquake, one of nature's largest disasters,
 results from the sudden release of huge amounts of energy in the Earth's crust,
 which creates seismic waves.
The hypocenter or focus is the point where the stored strain energy is first released
and the earthquake rupture begins.
Most hypocenters are located at the border
between the crust and mantle or at the very upper parts of the mantle of the Earth \cite{vill}.
The majority of tectonic earthquakes originate at the so-called `ring of fire'
and the depth of the hypocenters does not exceed tens of kilometers.
The epicenter is the point on the Earth's surface that is directly above the hypocenter.

A seismic wave is a kind of inhomogeneous acoustic wave.
There are two types of seismic waves: body waves and surface waves.
 Primary(P) and Secondary(S) waves are body waves,
 and  Rayleigh(R) and Love(L) waves are surface waves.
A surface wave is generated when body waves arrive at the surface of the earth
and the epicenter is the main point of generation.
Surface waves travel slower than body waves
and the amplitudes decrease exponentially with the depth.
Surface waves travel about $1 \sim 3km/sec$  and vary greatly within the depth of a wavelength \cite{vill,udias}.
 The wavelengths of surface waves are of the order of $100m$  and the frequencies are below  $30Hz$,
nearly below the audible frequency.
The amplitude of surface waves decays more slowly than that of body waves.
Those are of low frequencies of long durations and of large
amplitudes, which produce the most destruction causing serious
hazards to life and property.

Conventional earthquakeproofing methods of structural and geotechnical engineering
are aimed to ensure that the buildings themselves do not collapse from swinging or vibration.
These methods are generally effective, but basically comprise a passive response to earthquakes.
Humans cannot control the seismic waves and are thus still not safe from earthquakes.
However, recently developed methods provide a new way to resist seismic waves using acoustic metamaterials.
 Metamaterials are man-made effectively homogeneous structures  with dimensions potentially much smaller than
  that of a wavelength \cite{pend,smith,chan}.
 The seismic design with metamaterials is, in principle,  a form of cloaking.
 There are two types of cloaking methods: deflecting and attenuating seismic waves.

Deflecting seismic waves is a traditional cloaking method suggested by Farhat et al \cite{farh1,farh2,farh3}.
 They proposed a theoretical design of bending waves propagated in isotropic heterogeneous thin plates.
 Their design is a plate of concentric rings arranged from stiffest at the high refractive index
 to most flexible at the low refractive index from the outer ring to the innermost ring.
 The waves bend away from the foundation of the building as they pass through the plate.
 The cloaked seismic waves, however, are still destructive to buildings behind the cloaked region.
This does not solve the problem, but rather jeopardizes the surrounding buildings.

 The attenuated method is an extended cloaking method that are previously introduced \cite{kim,mit,discover,kim2}.
We suggested a method to create an artificial shadow zone using metamaterials.
Metamaterials act as an attenuator by converting the seismic wave into an attenuated wave
by making use of the imaginary velocity of the stop-band of the wave.
This method protects not only the building that is surrounded by metamaterials,
but also all buildings behind the metamaterials.
In this paper we report the method including a computer simulation of the shadow zone
and calculation of the sound level when the seismic wave is converted into a sound.


The phase velocity of all mechanical waves follows the expression of the form:
$v$= \{(elastic property)/(inertial property)\}$^{1/2}$.
Although  seismic surface waves are not purely two dimensional,
  velocity is mainly dependent on the density,
$\rho$, and shear modulus, $E$, of the seismic medium \cite{kim}.
Shear modulus, $E$, is  defined by $\Delta P = E \Delta x / h$, where
$\Delta x$ is the horizontal shift and $h$ is the height of the object.
The negative shear modulus causes an axisymmetric deformation under
an opposite axisymmetric loading.

The negative shear modulus of elastic media has been realized several times
 by Helmholtz resonators or its applications \cite{lakes,wu2,zhou}.
The resonance of accumulated waves in the Helmholtz resonator
reacts against the applied pressure at some specific frequency ranges.
Then, the negative modulus is realized by passing the acoustic wave
through an array of Helmholtz resonators called  ``meta-boxes."
Therefore, the acoustic intensity decays at some resonant frequency ranges.
    Considering the structural loss, the general form of the effective shear modulus of elastic materials,
$E_{eff}$, is given similarly with the general form of the bulk modulus
\cite{fang,lee1,lee2,kim,cheng1,cheng2}.
\begin{equation}
E_{eff}^{-1}={E}^{-1}\left[ 1-\frac{F \omega_o^2}{\omega^2 -
\omega_o^2 + i \Gamma \omega} \right],
 \label{15}
\end{equation}
where $\omega_o$ is the resonance frequency,  $\Gamma$ is a loss by damping,
 and $F$ is a geometric factor \cite{ding}.
The effective shear modulus has a negative value at $1 < \omega/\omega_o <
\sqrt{1+F}$ when the loss is small.
 The negative range of the real part is the stop-band of the wave.

If the shear modulus becomes negative, then the velocity becomes imaginary,
as  does the refractive index, $n$,
or the inverse of the velocity as $n=v_o/v=v_o \sqrt{\rho /E}$, where $v_o$ is the background velocity.
It has imaginary  impedance of absorption: $ \rho v = \sqrt{\rho E}$.
Because the wave-vector,  $k = 2\pi n/\lambda$, becomes imaginary, too,
 the amplitude of the seismic wave is attenuated.
 It functions as a ``barrier" or ``attenuator" to the seismic waves.


The intensity of an earthquake or the earthquake magnitude, $M$,
is defined by comparing the two amplitudes in a logarithmic scale \cite{vill,udias}.
\begin{equation}
M=\log \frac{A}{A_o},
\label{2}
\end{equation}
where  $A$ is the maximum amplitude of the seismic wave
and $A_o$  is the maximum amplitude of the background vibration.
$A_o$ depends on the distance from the epicenter and the order of micrometers.
If $A$ is recorded at $100km$ from the epicenter and $A_o = 1\mu m$,
then the magnitude becomes Richter-scale or local magnitude $M_L$.

Here we focus on reducing the amplitude of the seismic wave using the properties of the metamaterials.
 The energy of the seismic waves is dissipated inside of the meta-boxes,
and the absorbed energy is converted into sound and heat \cite{kins}.
The seismic energy is converted into sound the same way breathing pressure
 makes a resonant sound in a musical instrument.
 We built an attenuator or an earthquakeproof barrier to a seismic wave
 by filling up many resonators under the ground around the building that we want to protect.
 The amplitude of the seismic wave that passes through the meta-boxes is reduced
 by the imaginary wave-vector at the frequency ranges of the negative modulus.

\begin{figure}
\resizebox{!}{0.23\textheight}{\includegraphics{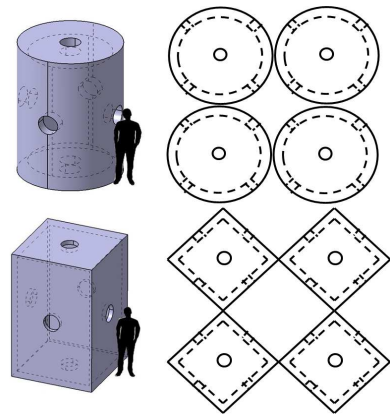}}
\caption{  Meta-boxes with 6 side holes and the positioning of the 4 meta-boxes.
The empty spaces among the boxes for passage between the holes
act as a large capacitor.
 }
 \label{fig2}
\end{figure}

The size of the meta-boxes can be estimated by analogy between electric circuits and mechanical pipes.
A pipe or tube with open ends corresponds to an inductor, and a closed end corresponds to a capacitor
\cite{kim}.
The resonance frequency of a Helmholtz resonator is \cite{ding,kins,bera}
\begin{equation}
\omega_o =\sqrt{\frac{S}{l' V}}v,
\label{17}
\end{equation}
where $S$ is the area of the cross-section,  $V$ is the volume, and $v$  is the background  acoustic velocity.
$l'$  is the effective length which is given by  $l' \simeq l + 0.85d$,
 where  $l$ is the length of the hole or thickness of the meta-box
 and  $d$ is the diameter of the hole \cite{bera}.

 An example of the design of the meta-boxes for the seismic frequency range is plotted in Fig. \ref{fig2}.
 It is a 6-hole cylinder or rectangular box. The diameter of the hole
 and the thickness of the cylinder is on the order of $0.3m$ or 1 foot.
 The volume inside could be  $10 \sim 100m^3$.
  The greater the number of holes, the higher the resonant frequency.
  The shape of the meta-box could be any form of a concrete box with a few side holes.
Because the meta-boxes are considerably smaller than the corresponding wavelengths,
 the array of the boxes behaves as a homogenized medium.
  Various types of resonators may cover various types of resonance frequencies of the seismic waves.

A seismic wave is an elastic wave that can be approximated by a sine wave.
Assuming a plane wave of wavelength $\lambda$ propagates in $x$-direction,
   the amplitude of the wave reduces exponentially \cite{kim}.
\begin{equation}
A e^{ikx} =  A e^{-2\pi |n| x/\lambda}.
 \label{20}
\end{equation}
We can rewrite Eq. (\ref{20}) with the definition of
the magnitude in Eq. (\ref{2}).
\begin{equation}
 10^{M_i}  e^{-2\pi |n| x/\lambda} =  10^{M_f},
\label{30}
\end{equation}
where $M_i$ and  $M_f$ are the seismic magnitude before and after entering the seismic barrier.
From Eq. (\ref{30}) we obtain the thickness of the seismic barrier $\Delta  x$ \cite{kim}.
\begin{equation}
\Delta  x = \frac{\ln 10}{2 \pi}\frac{\lambda \Delta M}{|n|}
=\frac{0.366 \lambda }{|n|}\Delta M,
 \label{35}
\end{equation}
where $\Delta M = M_i - M_f$.
Note that a high refractive index material is desirable for the narrow seismic barrier.
 The depth of the seismic barrier should be at least the same as the foundation of the building,
 but it is not necessary to be more than the wavelength of the surface waves.
 Magnitude 3 or lower earthquakes are almost imperceptible and rarely cause damage.
 On the other hand,  magnitude 6 or higher ones are very destructive,
and it is expensive to build buildings protected from these large earthquakes
using conventional earthquakeproof design.
The largest magnitude earthquakes on record were on the order of 9 \cite{vill}.
Most buildings in earthquake belts have a seismic design of $3 \le M \le 6$.
Therefore, $\Delta M = 3$  is the minimum magnitude required to effectively protect the buildings.

\begin{figure}
\resizebox{!}{0.18\textheight}{\includegraphics{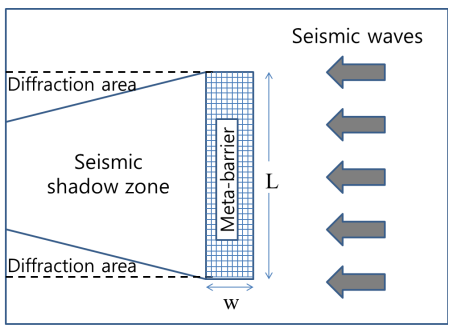}}
\caption{A design of $\Delta M =3$ meta-barrier made of many meta-boxes:
$w=60m$, $L=300m$.
The seismic shadow zone and its diffraction areas are shown. }
\label{fig6}
\end{figure}

We designed an artificial seismic shadow zone in Fig. \ref{fig6}
by constructing a meta-barrier which was composed of many meta-boxes.
The other side of the meta-barrier of the seismic wave is the seismic shadow zone
protecting all of the buildings.
The width of the barrier is on the order of the seismic wavelength,
but the length of the barrier should be much longer than the wavelength
 because there are diffraction waves from the both ends of the barrier.
Then, we may need an order of hundreds of thousand meta-boxes.
     If the relative refractive index of the meta-box is $|n|=2$
and the wavelength of the surface wave is  $\lambda=100m$,
a width of  $\Delta x \simeq 60m$ for $\Delta M=3$  resistance.
The shape of the barrier is neither circular  nor
 concentric rings like in the traditional clocking method \cite{farh1,farh2,farh3}.
 We have shown that water trench is not satisfactory for the attenuation of seismic wave \cite{kim}.

\begin{figure}
\resizebox{!}{0.35\textheight}{\includegraphics{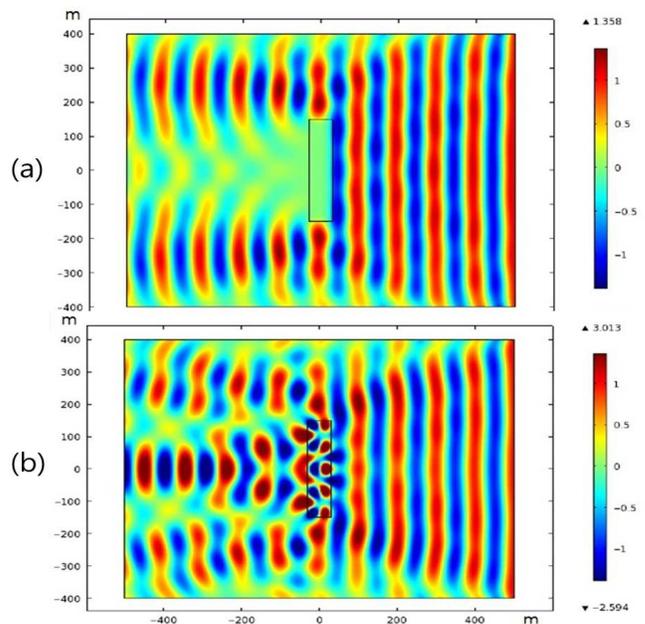}}
\caption{Pressure distribution by (a)a meta-barrier in Fig. \ref{fig6}
and (b)a normal barrier. freq.$=10Hz$.
 }
\label{fig7}
\end{figure}

The seismic shadow zone by a meta-barrier was displaced using  COMSOL multiphysics
and compared with a normal barrier.
We applied some typical material parameters.
 Density and modulus are $\rho=1.0 \times 10^3 kg/m^3$, $E=1.0GPa$ for the background.
The background could be rocks, water, or mixed.
We applied $\rho=2.0 \times 10^3 kg/m^3$ and $E=-0.50GPa$ for the meta-barrier,
and $E=0.50GPa$ with the same density for the normal barrier.
Note that the normal barrier has the refractive index 2
and matches the impedance $\rho v$ with the background.
We see from the Fig. \ref{fig7}(a) that the meta-barrier creates the seismic shadow zone
and diffraction areas.
On the other hand, the normal barrier in Fig. \ref{fig7}(b) can not stand from the seismic wave.


Converting mechanical energy into acoustic energy creates sound and heat.
It is simple to estimate the sound level.
The acoustic intensity or sound level in decibels(dB) is defined by \cite{kins,bera}
 \begin{equation}
\beta=10 \log \frac{I}{I_{ref}},
\label{45}
\end{equation}
where $I_{ref}$ is the reference sound intensity at the threshold of hearing
or $I_{ref}=1.0 \times 10^{-12} W/m^2$.
The intensity is proportional to the square of its amplitude.
Comparing Eq. (\ref{2}) with (\ref{45}), we can roughly estimate the sound level in decibels.
\begin{equation}
\beta \simeq 20 \log \frac{A}{A_{ref}} \simeq 20 M,
\label{47}
\end{equation}
where  $A_{ref}$ is the reference amplitude and $A$ is the root-mean-square of the amplitude
of the seismic wave.
Note that $M$ is the magnitude at the barrier and is different from Richter's scale $M_L$.

The sound level in Eq. (\ref{47}) can be rewritten with Richter's scale $M_L$.
The intensity of a three-dimensional wave decreases with $1/r^2$,
and that of a two-dimensional wave decreases with $1/r$, where $r$ is the distance from the source.
A seismic wave is a combination of these two.
Then, the magnitude $M$ is written as \cite{vill,udias}
\begin{equation}
M = a-b\log D,
\label{49}
\end{equation}
where $D$ is the distance from the epicenter, and $a$, $b$ are empirical positive constants.
 $b$ is the power of the distance and order of 1.5.
Because Richter's scale $M_L$ is defined at $D=100 km$,
 Eq. (\ref{49}) is written as $M_L=a-b\log 100$.
From the relation between $M$ and $M_L$, the sound level in Eq. (\ref{47})
is expressed with $M_L$ instead of $M$.
\begin{equation}
\beta \simeq 20 \left( M_L + b\log\frac{100}{D} \right),
\label{51}
\end{equation}
where $D$ is measured in $km$.
The dominant factor is not the distance $D$, but the magnitude $M_L$.
If $M_L =6$ and $D=100km$, then $\beta \simeq 120$dB,
 which corresponds to a siren or rock concert.
 However, the sound level is actually lower than this
because some part of the seismic energy is converted into heat by friction.
The frequency will be near infrasonic and barely audible.


We developed a different kind of cloaking method for the attenuation of seismic waves.
The method produces an artificial seismic shadow zone that abolishes
 the seismic wave by changing the seismic wave-vector to imaginary using meta-boxes.
We designed a meta-barrier and tested it by a computer simulation.
This method does not just add another seismic system to a building,
  but rather constructs an earthquakeproof barrier around the building to be protected.
   This barrier is a kind of waveguide that exponentially reduces the amplitude of the
   dangerous seismic waves.
The seismic energy in the barrier is converted into a very low frequency sound.

This method differs from the traditional cloaking method of deflecting seismic waves.
The traditional method sacrifices the buildings behind the meta-regions.
In addition, the seismic barrier method has several advantages.
The building we want to protect does not have to be altered, and
the seismic range of the building can be upgraded or downgraded by adjusting the width of the barrier.
This method will be effective for  buildings that are not well-equipped
with conventional seismic designs.
The biggest advantage is that all buildings behind the barrier will be protected,
 which decreases the cost  of seismic protection.
 This method is applicable for social overhead capitals, such as power plants, dams, airports,
nuclear reactors, oil refineries , long-span bridges,
 express rail-roads, historical monuments, among others.


This research was supported by  Basic Science Research Program through
the National Research Foundation of Korea(NRF)
funded by the Ministry of Education, Science and Technology(2012-0001425).


\end{document}